\documentclass[a4paper,12pt]{article} 
\usepackage{graphicx} 
\usepackage{graphics} 
\usepackage{amsmath} 
\usepackage{amsfonts}  
\usepackage{amssymb} 
\usepackage{bm}

\usepackage[colorlinks=true, pdfstartview=FitV, linkcolor=blue,  citecolor=blue, urlcolor=blue]{hyperref}  

\begin{document}

\title{Quantum mechanics as a statistical theory:\\ a short history and a worked example}
\author{
{Yves Pomeau$^1$ and   Martine Le Berre$^2$}
\\  
$^1$\small{Ladhyx, Ecole Polytechnique, Palaiseau, France}
\\ 
$^2$\small{Ismo, Universit\'e Paris-Saclay, Orsay , France}.
}

\date{\today }
\maketitle



\begin{abstract}
A major question in our understanding of the fabric of the world is where the randomness of some quantum phenomena comes from and how to represent it in a rational theory. The statistical interpretation of quantum mechanics made its way progressively since the early days of the theory. We summarize the main historical steps and then we outline how the randomness gains  to  be depicted by using  tools adapted to Markov processes. We consider a model system corresponding to experimental situations, namely a single two-level atom submitted to a monochromatic light triggering transitions from the ground to the excited state. After a short summary of present quantum approaches, we explain how a general "kinetic-like" Kolmogorov equation yields the statistical properties of the fluorescent light radiated by the atom which 
makes at once Rabi oscillations between the two states, and  random quantum jumps with photo-emission. As an exemple we give the probability distribution of the time intervals between two successive emitted photons by using the Kolmogorov equation. 
\end{abstract} 


\section{Introduction}
\label{intro}
 
The transition from Newtonian mechanics to quantum mechanics in the early years of the twentieth century has been a major step in the progress of our understanding of the world. This transition was more than a change of equations because it involved also a deep change in our understanding of the limits of human knowledge. It included from the very beginning a statistical interpretation of the theory. In other words quantum mechanics is not fully predictive and cannot be so. 

The introduction of statistical methods to describe Nature was not new of course. Originally  statistical concepts were introduced  to describe 
classically (not with quantum theory)
complex systems with many degrees of freedom like a volume of fluid including a very large number of molecules. These large systems  cannot be fully described and predicted since no human being has enough computational power to solve Newton's equations\footnote{We use reluctantly the word "equations" here because Newton never wrote down ordinary differential equation of classical mechanics in the modern sense. As well-known he solved dynamical problems by using elegant  geometrical methods instead of what we call now calculus. For instance the  solution of the two body problem with a general spherical potential was obtained by replacing integrals by area calculations.}
with the initial data (position and velocity) of too many particles: nowadays one cannot solve the classical equations of motion of more than a few thousand particles. In classical mechanics, another point makes difficult to predict distant future from the initial data. This problem occurs when a small disturbance or inaccuracy in the initial conditions is amplified in the course of time, a character linked to what is called the ergodicity properties of dynamical systems, which is very hard to prove for given systems. As far we are aware this is proved to be true \cite{reghard} only for systems of hard spheres making elastic collisions, and the proof is highly non-trivial. In these two examples (many particles and/or ergodicity of classical dynamics) the statistical method of analysis is just a way to describe systems  given the imperfect knowledge of the initial conditions and their overwhelming abundance. 

On the contrary quantum mechanics needs, from the very beginning, a statistical interpretation, a point that raised controversies. To many it seemed strange to postulate (see later for the precise meaning of this word in this context) a statistical interpretation of a theory that looks to be "deterministic" in the sense that the dynamical equations (Schr\"odinger or Dirac equations including the interaction with the electromagnetic field) look well posed with an unique solution for given initial data. This led and is still leading some to reject the statistical interpretation of quantum mechanics and to believe that it hides an inner structure of the world, yet to be discovered.  What is called "determinism" is however not as well defined as one could believe first. The  clearly defined {\it{mathematical}} meaning of this concept is based on the notion (seemingly first understood by Newton)
that for given initial data, like positions and velocities of particles moving in vacuo, there is a well defined future for a dynamical system obeying differential equations of finite order in time. A superficial view could give hope to predict the future if one has a a complete understanding of the initial data
because the equations of non relativistic quantum mechanics are mathematically "deterministic" and first order in time.
The fallacy of this idea comes from the word "complete". Because measurements of the initial conditions are made with quantum devices which have limited accuracy\cite{Heisen},  there is a   {\it{fundamental}} uncertainty in the initial conditions. This is central to our discussion  below concerning the emission of photons by an atom. Even if we have a perfect knowledge of the initial state of the atom and of  the measuring device, the instant of the emission of a photon cannot be predicted, although the evolution from perfectly accurate initial data is mathematically well defined.  

Many practicing scientists, including the authors, have no doubts on the validity of quantum mechanics and its so-called Copenhagen statistical interpretation. However in exact sciences like physics the so-to-say philosophical point of view is one thing but it is another one to put in practice general principles by writing equations and, hopefully by solving them. In Sec.\ref{shorthist} we outline the story of the statistical interpretation of quantum mechanics and explain why and how some points of this theory require to be used with great care in order to represent fairly experimental situations of the kind which motivated the founding fathers of this branch of physics.  
 Sec.\ref{A model} illustrate this remark.
We show in a non-trivial example, the fluorescence of a two-level atom, how one can relate explicitly  statistical concepts to the quantum phenomena one wants to describe. After a short outline of  previous quantum jump approaches based on  Lindblad equation (\ref{eq:lind}) for the density matrix, we show
in this example  how the  Markovian dynamics of the atomic state could be described by a Kolmogorov-like equation (\ref{eq:ko1}) which takes into account the  quantum jump events associated to the emission of photons in a statistical way.  
One of the goal of this paper is to draw attention to 
a theory of quantum phenomena using the apparatus
of statistical physics for the description of highly non equilibrium phenomena.

 \section{A short history of the statistical interpretation of quantum mechanics}
 \label{shorthist}
 
Let us recall the early developments of quantum mechanics in their relationship with our topic, with no claim of completeness. At the end of the nineteenth century three physical phenomena remained unexplained: the back-body spectrum, the existence of sharp lines in absorption and emission spectra of atoms and the photoelectric effect, showing a huge concentration of energy even in dim light beams. The challenge was to explain all of this.   

Quantum mechanics was the theory explaining those phenomena as well as many others, sometime with an unbelievable accuracy. Of course we think to quantum electrodynamics  phenomena, like the Lamb shift and the electron magnetic moment  corrections $(g-2)$ \footnote{The writing $(g-2)$ refer to what is called the anomalous magnetic moment of the electron. This magnetic moment is called $g$ and can be measured very accurately. Once the proper scalings are made, $g$ is equal to $2$ for a "pure electron" non interacting with the quantum fluctuations of vacuum, zero point fluctuations of the electromagnetic field and Dirac sea of electrons. The calculation of $(g-2)$ is done by perturbation with respect to the small coupling constant between the vacuum fluctuations and the electron. The dimensionless coupling constant, which is not that small, is the fine structure constant $\alpha =  \frac{e^2}{\hbar c}$. The Taylor expansion of $(g - 2)$ in powers of $\alpha $ has been done to the eighth order and agrees with the experimental value up to the twelveth digit, this making this theory the most accurately checked physical theory. The source of uncertainty is mostly in the accuracy with which the fine structure constant is known. To an outside observer it is also a bit curious that this expansion fits amazingly well the experiments whereas it is believed, after an argument  by Dyson  \cite{dyson} to have zero radius of
convergence.}.
 However some phenomena 
 related to the statistical interpretation of quantum mechanics took several years  to be interpreted,  as illustrated by the discovery of radioactive decay of heavy nuclei by emission of alpha particles (or alpha decay), which  is one of the first observations of quantum statistics. 
It
 was observed \cite{Rutherford} that
 a given proportion of the radioactive nuclei emits alpha particles per unit time so that the emission rate of a given sample and the population having not decayed decreases exponentially with time, the temporal time emission process following a Poisson Law.  This result is due to the fact that  the nuclei are uncorrelated, and that the emission process is random with a constant rate for each nucleus. At the time this was discovered, it was not recognized as a quantum effect, and no attempt was made to relate conceptually the randomness of the nuclear decay to other processes like the absorption of photons in the photoelectric effect discovered by Hertz. Nowadays, amazingly no entry of Wikipedia on alpha decay mentions  that it follows a Poisson law and has therefore  a fundamental statistical interpretation, and of course explains why it is so. 

Quantum mechanics started with the huge step(s) forward made by Planck when he explained \cite{Planck} the spectral distribution of the black body radiation. He did that in two steps. His first derivation was purely thermodynamical without modeling the interaction of atoms and light. It is at the urging of Boltzmann that Planck managed to derive the black-body spectrum from a more detailed physical model. As well known this model was based upon the idea that the energy of an oscillator instead of changing continuously, as in classical mechanics, changes by jumps of amplitude $ h \nu$, $h$ being a new physical constant (called Planck constant now) and $\nu$ the frequency of this oscillator. It is already evident that quantum mechanics, a least for simple situations, mixes classical concepts (the harmonic oscillator) and new physics, the one depending on Planck constant, a very small quantity, $6.62$ $10^{-34}$ J.s 
in units used for macroscopic physics. This constant has the physical dimension of a classical action. It is interesting to note that Planck had also the idea that the energy of a quantum oscillator  being in its ground state, is not zero but $h \nu/2$, something called now the zero-point energy. The existence of this zero-point energy has been spectacularly verified by measurements of the Casimir effect. Planck's derivation of the black-body spectrum was formally based on classical Gibbs-Boltzmann statistics and somehow skipped any details of a more complete theory. The next step was made by Einstein \cite{Einstein1}  when he showed that Planck's assumption explains also the photoelectric effect: the interaction of a light beam with matter is by quick jumps when an energy $ h \nu$ passes from the light beam to an illuminated atom. The balance of energy explains well also that, if the energy  $ h \nu$ is larger than the binding  energy of the electron with the atoms (a very coarse account of this complex phenomenon), some of it is transferred to the kinetic energy of this electron and so drives the observed  photocurrent. 

Again it cannot be said at this step of the developments that there was any obvious link between the nascent quantum theory and a fundamental statistical interpretation. This came in a short paper by Einstein in 1917, hundred years ago, where he derived \cite{Einstein2} Planck's black body spectrum by considering the  equilibrium steady state of a population of atoms interacting with light at equilibrium, namely a black-body radiation. He constrained the light spectrum by imposing that the exchanges between the two systems (atoms and light) keep each one at equilibrium but have a small effect by themselves on the equilibrium properties of the two systems, atom and radiation. This led Einstein to introduce three interaction coefficients, the one of spontaneous emission, the one of absorption and the one of induced emission. The induced emission and absorption are proportional to the intensity of light at the frequency difference between the two atomic states, whereas the spontaneous emission is independent of this intensity. Imposing that the ratio of populations of the excited and ground state is the Boltzmann factor, Einstein derived the Planck spectrum by assuming also that the spectrum becomes the Rayleigh spectrum at small frequencies. This particularly elegant derivation implied also that the transition from excited to ground state is a random process, not only on average over all the atoms, but also for every atom. Einstein clearly saw that such an assumption could not be explained by classical physics. All his life afterwards he remained reluctant to the idea of such a fundamental randomness. To make a long story short, it took ten more years until Dirac published   a paper \cite{Dirac}, when he was $27$ years old,  explaining how to derive Einstein's coefficients from the quantum dynamical equations for the coupled electromagnetic field and electrons in atoms. Somehow this closed, at least in some sense, the story, giving a fully rational basis for the calculation of a fundamental quantity of quantum mechanics. In particular it is significant to remark that Dirac's derivation explains why the return to the ground state by emission of a photon is a Poisson process with a rate derived from the "deterministic" quantum dynamics. Dirac takes a well defined initial condition for the full system: the atom is in an excited state whereas the electromagnetic (EM) field is itself in its ground state, with photons having the zero-point energy. Because of the coupling between the electrons of the atom and the EM field, this initial condition is a priori not an exact eigenstate of the full system. Following the general principles of quantum mechanics, the system atom+EM field has another state with the same energy, namely the emitted photon and the atom returned to the ground state. Therefore the mixing between the two states of equal energy is possible, and, once the interaction between the EM field and the atom is turned on, the amplitude of the "other state" (emitted photon + atomic ground state) grows from its  zero initial value. This amplitude grows at the beginning proportional to time. The coefficient of time is interpreted by Dirac as a rate of transition from one state to the other. When proving this point, Dirac very thorough analysis uses the fact that the quantum phases of the two states are different and random, which is fundamentally why the result has a statistical meaning. The randomness of the phase is there because the emitted photon goes away by carrying its own phase which becomes quickly uncorrelated to the one of the emitting atom, leading to an irreversible process. If, instead of this photo-emission process, one had considered  the case of an atom trapped  in a lossless cavity which is initially  in one of the  two quasi-degenerate states formed by \{atom in the ground state+1 photon, excited atom + zero photo\},  then  when starting from one of the two states,  the amplitude of probability of the system to be in the other state starts to grow,  then an oscillation takes place between the two states. In this case the  cavity send back the photon on the atom which re-absorbs it, as if they play an endless atomic squash game,  during which the two states  keep a correlated phase. In this case the oscillations are reversible.

After the early days of this grand history of the birth of quantum mechanics, a somewhat arcane field of knowledge had to be transformed into the matter of lectures for (usually advanced) students, a teaching that came rather late in countries like France, not before the late nineteen forties at the earliest. This teaching had to face the question of explaining why such a theory with seemingly well posed dynamical equations (Schr\"odinger and Dirac equations) had to have a kind of fundamental statistical interpretation. The explanation relied on what is often called the "reduction (or collapse) of the wave packet",  a somewhat obscure concept. As often pointed out it is unclear if it is a fundamental lemma, namely an independent law to be added to the corpus of assumptions of quantum mechanics, or if it is a consequence of the rest of the theory. The difficulty with this concept is that quantum mechanics conserves probability because the evolution equations are exactly unitary, the probability being understood as the modulus square of the wave function. In other terms, interpreting this square of the norm as a probability, the property that the evolution is unitary is equivalent to state that the total probability is conserved. Therefore any "reduction/collapse of the wave packet " seems to go against this fundamental property: if a measurement amounts to "reduce " the wave packet, by particularizing the evolution of the system to data coherent with this measurement, this leaves undetermined the fate of the other contributions to the state of the system, corresponding to other results of this measurement or to other "reductions of the wave packet", all contributing additively to the squared norm, equal to one {\it{before}} and {\it{after}} the measurement: the other contributions to the norm have to be somewhere to conserve the total norm, or total probability. Later Everett introduced \cite{everett} a convincing explanation compatible both with the idea of reduction of the wave packet and the constraint of unitarity of the evolution, or of conservation of the probability in the statistical interpretation. Everett's idea is that each outcome of a measurement 
defines the universe conditional to the observer
disconnected of other universes corresponding  to another observer (or another outcome of the measurement).  As said  Hawking ``all that one does is to calculate conditional probabilities``. The profound idea  of Everett makes everything consistent, at the price of introducing a direction of time. This direction of time plays the same role as the one introduced to explain the arrow of time of thermodynamics, it just represents the physical impossibility to reverse the history of a peculiar system. Said otherwise, the statistics introduced by quantum mechanics is there, in principle, to make averages over all universes  relative to a given observer.
As said above, because we are discussing something related to physics and not philosophy, there are consequences of this line of reasoning in the physical and mathematical picture of processes. This relies on definite equations for probability distributions, of which we shall give an example below.  

Everett's theory is seen sometimes as going against everyday experience of a single history of the universe and of every thinking individual in it. 
This raises an interesting issue which has been there forever: does human mind (not to use the word "consciousness" or even the non-scientific "soul")  behave according to the laws of physics or is there something special about it? There is no evidence that our brain does not behave according to the laws of physics.  For instance this behavior seems to be consistent with the conservation of energy and the increase of entropy (to name two important laws of physics). Once one admits that Everett's interpretation includes everything in each different universe, one has to admit too that thinking and feeling people have multiple lives, each one in a specific universe, even though each one believes he lives an unique life, without bifurcation, at least in the physical sense. Hard to believe perhaps but nothing goes against it!  

\section{A model physical problem}
\label{A model} 

Here we consider the emission of radiation by a single atom, a subject which has been largely studied in quantum mechanics but merits further considerations in  our opinion. We hope that this beautiful example arouses the interest in the statistical community.

 The spontaneous emission of a photon by an atom initially in an excited state was long ago considered by Einstein and by Dirac. This introduced a quantum process with a fundamental randomness showing up in the Markovian point process defined by the photo-emission times. In particular  the  time intervals between successive emissions are independent and identical random variables. 
 Thanks to the progress of experimental atomic physics the discrete character of the emission times have been observed  in a slightly more complex situation proposed by Dehmelt \cite{Dehmelt}, where the  fluorescence of  a single three-level atom is intermittent because the atom is, so to speak,  maintained in one of the two excited states for a long time. 
 As in the two-level atom case, this atom is subject to two phenomena, first under the effect of a pump field it tends to make oscillations between the ground and the excited state. Once in this excited state it can  also jump spontaneously to its ground state by emitting a photon. The process goes on forever provided the pump field is on.

\subsection{ Lindblad equation }
\label{sec:lindblad}
Let us briefly discuss previous  descriptions of fluorescence 
 which have been used to model the emission of photons by a single  two-level (and three level) pumped atom. A first group of quantum treatments treats the laser field as classical. Starting  from Schr\"odinger or Heisenberg equations,  it leads to the Heisenberg-Langevin equations when using the Markovian  (or short memory) approximation. The coupled equations for the atomic operators $S_{+},S_{-}, S_{z}$   combine dissipation, excitation and  fluctuation terms.  The latter term being hard to handle,  can be cancelled by making an average over the vacuum field. It gives the optical Bloch equations for the expectation values  $<S_{+,-,z,}(t)>$, or equivalently to  a Lindblad-like equation \cite{lind} for the $2$x$2$ atomic density matrix, of the form
\begin{equation}
 \dot{\rho}   = -i[H, \rho] - (A A^{\dagger} \rho  + \rho A A^{\dagger}) + 2 A  \rho A^{\dagger} 
  \textrm{,} 
 \label{eq:lind}
\end{equation}
where  $H$ is the Hamiltonian for  an atom interacting with the classical  pump laser, and $A=\gamma <S_{+}> $ is the atomic raising operator, times $\gamma$, the Einstein coefficient associated to the spontaneous emission. This equation conserves the trace of the density operator, namely the sum of the atomic populations,
 but forbids  to describe individual quantum trajectories and then to derive the single-particle properties.

After a workshop in Copenhague \cite{nordita}, quantum jump approaches  started to develop, in particular with the work of the LKB group in Paris \cite{lkb} who described  the fluorescence as a  cascading process downwards lower energy states  of the dressed atom which emits photons at each step.  To get this colorful description, the authors start from Lindblad equation (\ref{eq:lind}) 
for the 
 density operator $\rho(t)$ of the quantum system \{atom+pump field\}. 
In this case (\ref{eq:lind}) is  derived from the Schr\"odinger equation as above but one has to  make a a lot of supplementary hypothesis (commutation of operators, short memory of the coupling between the system and the environment and so on) before taking the  trace over the environment. The density matrix is  now infinite dimensional, because the states are labelled by two quantum numbers, ($g,e$ for the atom and $n$ for the laser photon number).  In \cite{lkb}  the authors  were able to derive an expression for the  ''delay function'' between two emitted photons, at the price of  focusing on a single step of the cascading process,  that reduces the
 infinite set of coupled equations to only two.  This simplification amounts to drop the accreting term $2 A  \rho A^{\dagger} $  in  
 (\ref{eq:lind}), that  
leads to a nonconservative evolution of the populations, even in the absence of photo-emission.  Such a  non conservative  dynamics, 
was largely  used from this time, and justified by the argument that  
'' not only  does the detection of a photon lead to an increase of the information, but the failure to detect  a photon does as well '' \cite{plenio}. 

Because this argument is actually an assumption which seems questionable, we think it is necessary  to describe fluorescence as a process which keeps the unitary character of the evolution and  take account of jump events in a statistical way. This is the basis of the theory sketched just below, based on what we call a Kolmogorov equation.

   \subsection{Toward of full statistical theory of the emission process}
\label{toward}

The statistical  theory we propose follows a statement  \cite{kolmo} on the axiomatics of random variables  which has applications in fields of science that have no relation to the concepts of random events and of probability in the precise meaning of these words. 
We shall  outline the principles of a statistical treatment able to describe both  the emission of photons and  the optical Rabi oscillations in the case of a  single pumped two-level atom, detailed in \cite{statphys}, then we shall explain 
 how to derive the probability distribution of the time intervals between two  successive photo-emission events. This was based upon the property that, in such an interval the atom does make unhindered Rabi oscillations which are interrupted by the emission of a photon, a phenomenon seen as instantaneous. This is of course one basic feature of a Markov process, because we consider  quick jumps occurring at random with a probability depending on the state of the system and, possibly, on the absolute time. 
For such a phenomenon 
the Kolmogorov equation seems  the right tool to describe the  state of an atom whose wave-function is of the form,
  \begin{equation}
  \Psi_{at}(t) = \left(\cos(\theta(t))  \vert g>  + i e^{i \omega t} \sin(\theta(t))  \vert e> \right) e^{i \varphi} 
 \textrm{,} 
 \label{eq:V}
\end{equation}
where $\dot \theta = \Omega/2 $ between two jumps,  $\Omega$ being the Rabi frequency. The Kolmogorov equation
 deals explicitly with the probability distribution $p(\theta,t)$ for the atomic state, here  indexed by a single variable $\theta$, and
has a built-in conservation law of the total probability at any time, a serious advantage with respect to the quantum treatments using truncated Lindblad equation with nonconservative interaction Hamiltonian in the inter-emission intervals. 
The derivation of the evolution equation of  $p(\theta, t)$ is explained in \cite{statphys}. This equation includes on its left-hand side a streaming term representing the Rabi oscillation, whereas the right-hand side includes two terms, both representing the effect of random emission of photons leading to a return to the ground state. This right-hand side has the familiar structure of the Kolmogorov equations for Markov processes with a gain and a loss term. The final result for a resonant pump field\footnote{For a detuned laser field, the Kolmogorov equation keeps a form similar to (\ref{eq:ko1}) \cite{statphys} with different coefficients.} is,     
\begin{equation}
\frac{{\partial {p}}}{\partial {t}} + \frac{\Omega}{2} \frac{{\partial {p}}}{\partial {\theta}} = \gamma \left(  \delta(\sin\theta)  \left(  \int_{-\pi/2 }^{\pi/2} \mathrm{d}{\theta'}\ p(\theta', t)  \sin^2\theta' \right) - p(\theta, t)  \sin^2\theta  \right)
 \textrm{.} 
\label{eq:ko1}
\end{equation}
Introducing a probability distribution depending on a continuous variable,
$\theta$ here, is a way to take into account all possible trajectories
emanating from the emission  of a single photon, with a new value of the
number of photons  radiated in any direction at each quantum jump.
Average values of a time depending quantity which depends on $\theta$ can
be calculated   via the probability distribution  $p(\theta,t)$ which is a $\pi$-periodic function with a finite jump at $\theta=0$
 but smooth elsewhere.
Such procedure allows to deal correctly with the infinite number of
possible trajectories,  since Boltzmann's genius lies precisely in transforming the classical statistical theory based on unknown initial conditions into statistics for an ensemble of indeterminate trajectories help to the ergodic hypothesis. 

Because it is linear this equation can be solved in Laplace transform, but the general solution in time requires the inversion of a Laplace transform which can be done only formally. 
There are two constraints, (\textit{i}) the probability $p(\theta, t)  $ is positive or zero and (\textit{ii})
the  total probability  $ \int_{-\pi/2 }^{\pi/2} \mathrm{d}{\theta}\ p(\theta, t) $ is unity  at any time, that reflects the unitary evolution of the atomic state (the integral of the square of the wave function is constant and equal to one).  It is relatively easy to check that they are fulfilled, since $ \int_{-\pi/2 }^{\pi/2} \mathrm{d}{\theta}\ p(\theta, t) $ is constant and $p(\theta, t) \geq 0$ at any positive time if $p(\theta, 0) \geq 0$.
Solutions  in various limits are derived in \cite{statphys}. 
The factors $\sin^2\theta $ on the right-hand side are there to take into account that a quantum jump occurs only if the atom is in the excited state, which has probability  $\sin^2\theta$. The negative term on the right-hand side is the loss term representing the decrease of the amplitude of the excited state by jumps to the ground state, whereas the positive one is for the increase of amplitude of the ground state when a jump takes place. 

 The probabilities  for the atom to be in the excited or in the ground state are respectively, 
\begin{equation}
 \rho_{1} (t) =  \int_{-\pi/2 }^{\pi/2} \mathrm{d}{\theta'}\ p(\theta', t)  \sin^2\theta' \textrm{.}
\label{eq:a1}
\end{equation}
 and 
\begin{equation}
\rho_{0} (t) =  \int_{-\pi/2 }^{\pi/2} \mathrm{d}{\theta'}\ p(\theta', t)  \cos^2\theta' \textrm{.} 
\label{eq:ao}
\end{equation}

 Their sum is one, as it should be, if $p(\theta, t)$ is normalized to one. From (\ref{eq:ko1}) one can derive an equation for the time dependence of $ \rho_{1} (t)$ and $ \rho_{0} (t)$ by multiplying (\ref{eq:ko1}) by  $\sin^2\theta$ and by $\cos^2\theta$ and integrating the result over $\theta$. It gives,  
 \begin{equation}
 \dot{\rho}_1 = - \frac{\Omega}{2}  \int_{-\pi/2 }^{\pi/2} \mathrm{d}{\theta'}\ \sin^2\theta'\frac{{\partial {p}}}{\partial {\theta}} -  \gamma \left( \int_{-\pi/2 }^{\pi/2} \mathrm{d}{\theta'}\ p(\theta', t) \sin^4\theta' \right)
 \textrm{,} 
\label{eq:ko1.2}
\end{equation}
and
\begin{equation}
 \dot{\rho}_0 = - \frac{\Omega}{2}  \int_{-\pi/2 }^{\pi/2} \mathrm{d}{\theta'}\ \cos^2\theta'\frac{{\partial {p}}}{\partial {\theta}} +  \gamma \left( \int_{-\pi/2 }^{\pi/2} \mathrm{d}{\theta'}\ p(\theta', t) \sin^4\theta' \right)
 \textrm{.} 
\label{eq:ko1.1}
\end{equation}

In the r.h.s of the rate equations (\ref{eq:ko1.2}-\ref{eq:ko1.1}) associated to the populations of the two levels, the first term, proportional to the Rabi frequency $\Omega$, describes the effect of the Rabi oscillations, whereas the second term, proportional to $\gamma$, displays the effect of the quantum jumps responsible for the photo-emission. Because  
$p(\theta,t)$  includes  both the fluctuations due to the quantum jumps and  the streaming term, the right hand side of (\ref{eq:ko1.2})-(\ref{eq:ko1.1})  represents the new history beginning at each step.
After integration by parts (\ref{eq:ko1.2}-\ref{eq:ko1.1}) become, 
 \begin{equation}
 \dot{\rho}_1 (t)=   - \dot{\rho}_0 (t)= \int_{-\pi/2 }^{\pi/2}   \mathrm{d}{\theta}\; p(\theta, t)\; (\frac{\Omega}{2}\sin 2\theta -  \gamma \sin^4\theta)
 \textrm{.} 
\label{eq:lam}
\end{equation}

Note that the set of equations (\ref{eq:ko1.2}-\ref{eq:ko1.1}), or (\ref{eq:lam}),  is not closed. It {\it{cannot}} be mapped into equations for $ \rho_{1} (t)$  and $ \rho_{0} (t)$ only because their right-hand sides depend on higher momenta of the probability distribution $p(\theta, t)$, momenta that cannot be derived from the knowledge of $ \rho_{1} (t)$  and $ \rho_{0} (t)$. The unclosed form of  (\ref{eq:ko1.2})-(\ref{eq:ko1.1})  is a rather common situation. To name a few cases, the BBGKY hierarchy of non-equilibrium statistical physics makes an infinite set of coupled equations for the distribution functions of systems of interacting (classical) particles  \cite{BBGKY} where the evolution of the one-body distribution depends explicitly on the two-body distribution, that depends itself on the three-body distribution, etc. In the theory of fully developed turbulence the average value of the velocity depends on the average value of the two-point correlation of the velocity fluctuations, depending itself on the three-points correlations, etc.
Fortunately, one can  solve the Kolmogorov equation (\ref{eq:ko1}) via an implicit integral equation \cite{statphys},  then there is generally no need to manipulate an infinite hierarchy of equations as in those examples. 

In the present case one can say, following Everett,  that
  the probability distribution $p(\theta, t)$ allows to make averages over the states of the atom in different worlds, each being labeled by a value of ${\theta}$ at a given time $t$. 
 As written above, physical phenomena like the observation of a quantum state decay measured by emission of a photon, is relative to the measurement apparatus which takes place  in the universe associated to the observer. At every emission of photon a new history begins,
 represented by the right-hand side of (\ref{eq:ko1.1}).  
 In summary the creation of new universes at each step  defines a Markov process, which can be described by a Kolmogorov statistical picture, and  cannot be considered as a deterministic process depending in a simple way on averaged quantities like population values.

 To illustrate how one can use Kolmogorov equation, we  derive the time dependent probability of photo-emission by  a single atom, first without any pump field, then in presence of a resonant laser.
 We consider first an isolated atom initially in pure state $ \Psi_{at}(0) $ given by (\ref{eq:V}) with $\theta(0)=\theta_{0}$. 
 We search for the evolution of the probability $\rho_{1}(t)= \int_{-\pi/2 }^{\pi/2} \mathrm{d}{\theta}\ p(\theta, t)  \sin^2\theta \textrm{.}$ that the atom is in excited state at time $t$.  From (\ref{eq:ko1})  with $\Omega=0$ (no pump), we deduce that the emission of a photon occurs randomly in time with a rate,
 \begin{equation}
\dot{\rho_{1}}= \gamma \sin^{2}\theta (t) \rho_{1}(t) 
  \textrm{.} 
 \label{eq:rate}
\end{equation}
Once the atom jumps to its ground state it cannot emit another photon, then 
 the emission of a photon, if recorded, is a way to measure the state of the atom.
The solution of (\ref{eq:rate}) leads to the population of the excited state $\rho_{1}(t)= \sin^{2} \theta_{0} e^{-\gamma  \sin^{2}\theta_{0} \; t}$   when taking into account the initial condition, and the photo-emission rate is,  
 \begin{equation}
\dot{\rho_{1}}(t)= -\gamma\sin^{4} \theta_{0} e^{-\gamma  \sin^{2}\theta_{0} \; t}
 \textrm{.} 
 \label{eq:p1}
\end{equation}
 The probability of photo-emission  in the interval $(0,\infty)$ is   the integral of $\dot{\rho_{1}}$
   \begin{equation}
 \int_{0}^{\infty} \gamma \sin^{4}\theta_{0} \;  e^{-\gamma  \sin^{2}\theta_{0} \; t} dt=\sin^{2}\theta_{0}
   \textrm{,} 
 \label{eq:normp1}
\end{equation}
which means that the final state of the coupled system atom+emitted photon field is
   \begin{equation}
\Psi(\infty) = \sin \theta_{0}  \vert g,1>  +  e^{i \phi'} \cos \theta_{0}  \vert g,0> 
   \textrm{,} 
 \label{eq:psif}
\end{equation}
 where  the indices (1,0) correspond to  one and zero photon state respectively.The relation (\ref{eq:normp1}) means that if we consider $N$ atoms initially prepared in a given pure state with $\theta(0)=\theta_{0}$, namely with total energy  $N \sin^{2}\theta_{0} \hbar \omega$, we get at infinite time, $N$ atoms in the ground state and $N \sin^{2}\theta_{0}$ photons of individual energy $\hbar \omega$.  In the final state  only a fraction of  them, $N \sin^{2} \theta_{0}$, did  jump from the excited state to the ground state, with the emission of a photon, the other part, $N \cos^{2} \theta_{0}$, simply stayed  in the ground state\footnote{We thank  C. Cohen-Tannoudji, J. Dalibard and S. Reynaud for a stimulating discussion which was at the origin of the above derivation.}.
   \begin{figure}
\centerline{
(a) \includegraphics[height=1.5in]{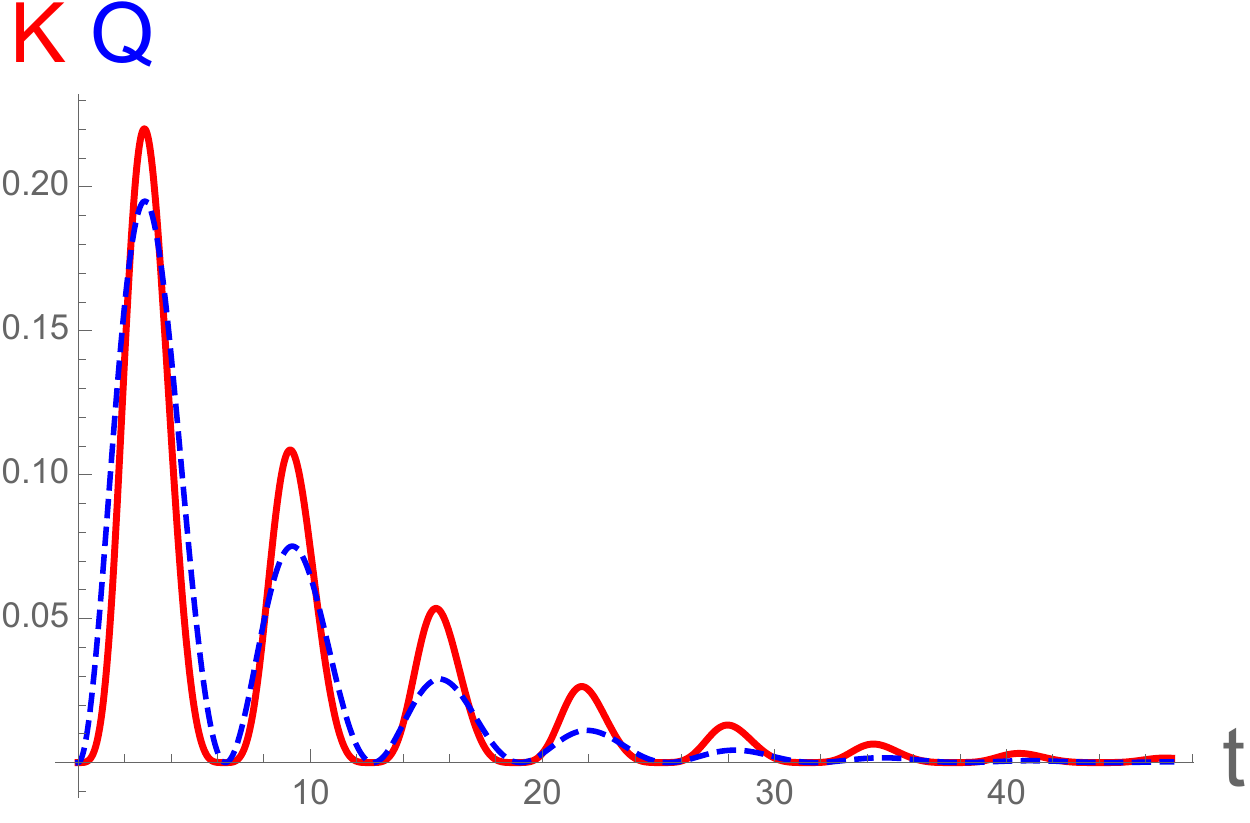}
(b) \includegraphics[height=1.5in]{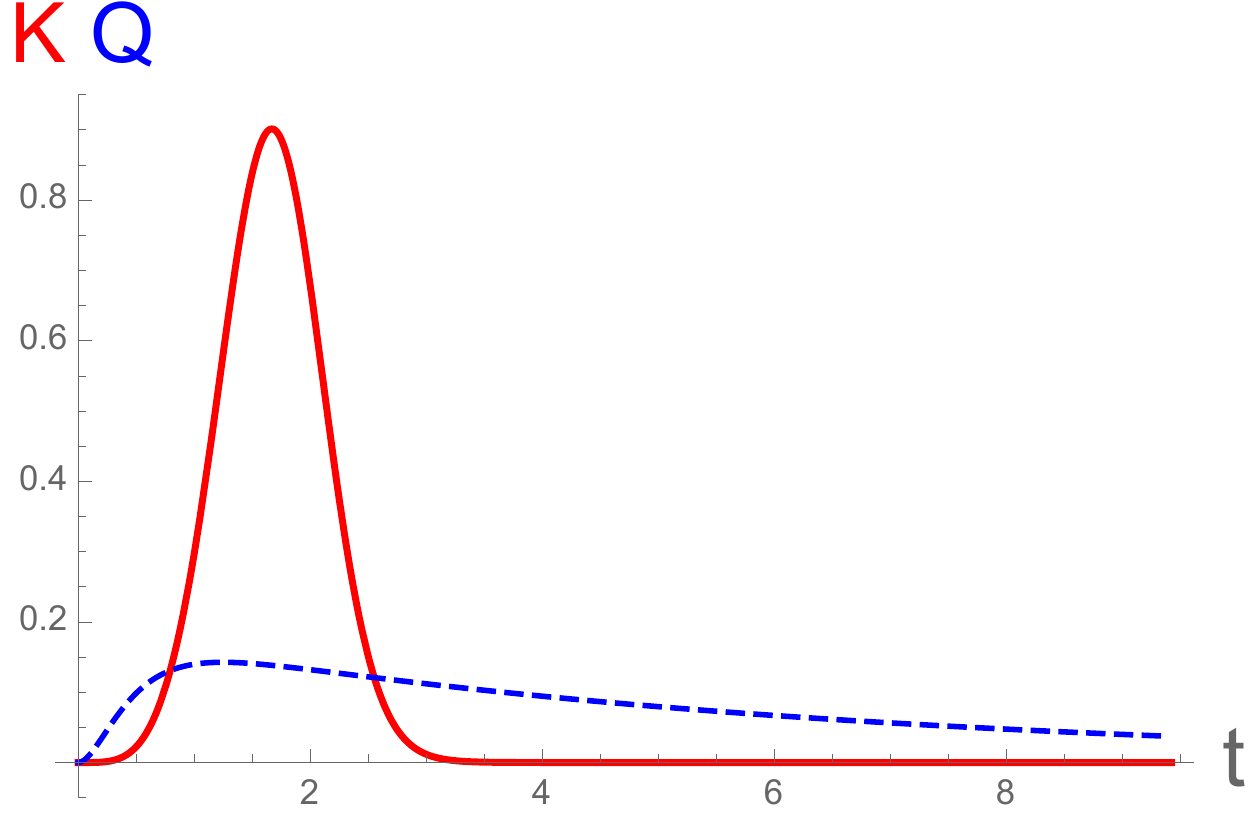}
  }
\caption{\small{Inter-emission time distribution $\ell(t)$ in two opposite cases, (a) for weak  (b) for strong dissipative rate (with the respect to the Rabi frequency). The solid red curves are for our Kolmogorov statistical theory, equation  (\ref{eq:ell}).  The blue  dashed curves display the delay function  deduced in \cite{lkb} for same values of $\Omega/\gamma$ which are equal to  $3.33$ in (a) and $1/6$ in (b)}.
}
\label{fig:tau}
\end{figure}

In the case of an atom submitted to a resonant pump field,  the atom will emit  photons at random times forming a point process. Here we assume that the process is Markovian, but more generally any process with  time-dependent history, is completely characterized by its conditional intensity function $\lambda(t\vert \mathcal{H}_{t})$, the density of points at time $t$, 
where 
$ \mathcal{H}_{t}$ is the  history of the emission activity up to time $t$, and the time interval distribution is given by the relation,
$\ell(\tau)=   \lambda(\tau \vert \mathcal{H}_{\tau})   e^{-\int_{0}^{\tau}  \lambda(t\vert \mathcal{H}_{t})dt  } $.
In  the present Markovian case the conditional intensity of the point process, which is
the probability of emission of a photon  at time $t$, only depends on the value of $\theta$ at this time, therefore one has simply $\ell(\tau)=   \lambda(\tau)   e^{-\int_{0}^{\tau}  \lambda(t)dt } $. From (\ref{eq:ko1.2}) we deduce $\lambda(t)=\gamma  \sin^{4}\theta(t)$.
In between two successive emission times the  atom undergoes Rabi oscillations, with $\theta(t)=\Omega t /2$, assuming a photon is emitted at time $t=0$. Therefore 
 the inter-emission time distribution  for an  atom  driven by a resonant pump is given by the expression\footnote{We take the opportunity of this publication to give the right expression of the inter-emission time probability, labelled as $K(t)$ in \cite{statphys}  where  $ \sin^{2}(\frac{\Omega}{2} \tau)$  should be changed into $ \sin^{4}(\frac{\Omega}{2} \tau)$ in the expression of the conditional intensity of the photo-emission point process},
   \begin{equation}
\ell(\tau)=  \gamma  \sin^{4}(\frac{\Omega}{2} \tau)\;\;  e^{-\gamma\int_{0}^{\tau}   \sin^{4}(\frac{\Omega}{2} t) }   
   \textrm{,} 
 \label{eq:ell}
\end{equation}
which gives  $\int_{0}^{\infty} {\ell(\tau) d\tau}=1$, as expected.  The result is  shown in Figs.\ref{fig:tau} in the two opposite limits of  large and small  values of the ratio $ \Omega/\gamma$, and compared to  the delay function derived in \cite{lkb} (which has not the standard form expected for a Markovian process).
For the case of strong input field, $\Omega >  \gamma$  the two methods agree approximately, see Fig.a. But  they differ noticeably for the opposite case  shown in Fig.b. For weak laser intensity (or strong damping) the Kolmogorov derivation gives a mean delay between successive photons of order $ \tau_{K} =  (\Omega^{4}\gamma)^{-1/5}$, which decreases  slowly as the damping rate $\gamma$ increases,  that seems reasonable. In the same limit the dressed atom method leads to 
$ \tau_{Q} =  \gamma/\Omega^{2}$, a time scale much longer than the inverse of $\gamma$, and increasing with the damping rate, a  strange result\footnote{ in the 1986 paper of Cohen-Tannoudji and Dalibard the authors interpret $ 1/\tau_{Q}$  as the width of the ground state induced by the pump laser.  We must also notice that the average number of radiated photons per unit time deduced from the Bloch equations is also of the order  $\gamma/\Omega^{2}$ in this limit.}.

\section{Summary and conclusions}
\label{summary}

The purpose of this paper was to show first how the view of quantum mechanics grew from the very beginning as a statistical theory,  and how things got clarified by Everett's bold idea of multi-universes. We felt also that it was not sufficient to discuss those questions abstractly as points of metaphysics\footnote{Metaphysics is here understood  as coming just after physics, in its original meaning by Aristotle, although the word "metaphysics" is not by Aristotle.} but needs to be debated as point of physics. This was demonstrated on a model problem with a non trivial "solution", namely a model where the statistical analysis needs to be done very carefully even though its mathematics is actually fairly simple. This model has also the interest to be connected with the problems raised first by the founding fathers focused on the interaction of matter and light. We thought it was instructive to show how the general concepts of quantum mechanics as a statistical theory work concretely in a given case. By concretely we mean in a probabilistic mathematical framework using probability distributions and their evolution equation. We hope that this discussion of a specific model brings more light on this difficult subject than a more abstract discussion.

To take a wider view of the problem, it is of interest to recall that {\emph{classical}} mechanics took a long time to become a theory not requiring a specific philosophical approach. For instance in the middle of the eighteen century a fierce debate took place between the deterministic view of Newtonian mechanics, where the initial conditions determine the future, and the Maupertuis view where the evolution is dictated by the minimization of an action integral with boundary conditions at the two ends of the time interval. Nowadays we know that the two pictures of classical mechanics are equivalent, but "philosophically" there is an obvious difference between these two interpretations of classical mechanics (via differential equation or via a minimization principle with ends fixed). So it is not that surprising, seen on the long term evolution of Science, to observe that after a century of development still some difficulties and misunderstanding remain in the interpretation of quantum mechanics. This work tried to contribute, not only to the exposition of the topic but also to put forward the idea that the treatment of quantum mechanics as a statistical theory is not such a trivial matter and that this should be done carefully, somehow by using at least implicitly the general principles of statistical theory, which could be seen perhaps as the  "metaphysical side" of this physics.

'footnote

\section*{acknowledgments}
The authors greatly acknowledge Jean Ginibre for interesting and fruitful discussions

{}

\end{document}